\documentstyle[preprint,revtex,eqsecnum]{aps}

\newcommand{\f}{\frac}
\newcommand{\she}{\hat s}
\newcommand{\lp}{L^+}
\newcommand{\lm}{L^-}
\newcommand{\ra}{\rightarrow}
\newcommand{\tw}{\theta _{\small W}}
\newcommand{\zp}{((\she-m_{\small Z}^2)^2+\Gamma_{\small_Z}^2m_{\small Z}^2)}
\newcommand{\s}{\sigma}
\newcommand{\wpr}{((\she-m_{\small W}^2)^2+\Gamma_{\small_W}^2m_{\small W}^2)}
\newcommand{\hp}{((\she-m_{\small H}^2)^2+\Gamma_{\small_H}^2m_{\small H}^2)}

\newcommand{\qll}{q \bar q\ra \lp \lm }

\begin{document}

\preprint{IFP-439-UNC}
\preprint{TRI-PP-92-97}
\preprint{WM-92-111}
\preprint{October 1992}
\begin{title}
SEARCH FOR HEAVY LEPTONS AT HADRON COLLIDERS
\end{title}
\author{Paul H. Frampton$^{(a)}$, Daniel Ng$^{(b)}$,\\
 Marc Sher$^{(c)}$ and Yao Yuan$^{(c)}$}
\begin{instit}
$^{(a)}$ Institute of Field Physics, Department of Physics and Astronomy
\\ University of North Carolina,  Chapel Hill, NC 27599-3255\\
$^{(b)}$ TRIUMF, 4004 Wesbrook Mall, Vancouver, B.C. V6T 2A3, Canada\\
$^{(c)}$ Department of Physics, College of William and Mary, Williamsburg,
VA 23187
\end{instit}
\vspace{-0.5cm}
\begin{abstract}

Four models are considered which contain heavy leptons beyond the
three families of the standard model. Two are fourth-generation
extensions of the standard model in which the right-handed heavy leptons
are either isosinglets or in an isodoublet; the other two are
motivated by the aspon
model of CP violation. In all these models,
the heavy neutrino can either be
heavier than, or comparable in mass to, the charged lepton leading
to the possibility that the charged lepton is very long-lived.
Production cross section and signatures for the heavy
leptons are computed for the SSC and LHC.
\end{abstract}
\newpage

\section{introduction}
Since the accurate measurement of the parameters of $Z^0$
decay\cite{r1}, it has been known that there exist only
three light
neutrinos, $\nu_ e,\ \nu _u \ $and $\nu _{\tau},\ $coupling
to the $Z^0$ in the manner prescribed
by the standard model. The simplest supposition
is then that the lepton sector comprises these three light
neutrinos and their charged counterparts, $e,\ \mu $
and $\tau $. However, it is quite possible
that heavy leptons exist. Such heavy leptons, which shall be
 designated as
$ L$ and $N $, for the charged and neutral varieties respectively,
will be a target of investigation
at the next generation of particle colliders, most notably
the Superconducting Super Collider(SSC) and the Large
Hadron Collider(LHC). In this paper we
 specify four simple models which contain such heavy
leptons and calculate their production cross sections
at the SSC and LHC. The first two models are fourth-generation
models where the right-handed $L$ and $N $
are doublets and singlets respectively
under electroweak $SU(2)$. The third and fourth models are
inspired by the aspon model\cite{r2} of CP
violation.

Many analyses of heavy lepton production have previously
been done\cite{ra1}.
Our work differs from Ref. \cite{ra1} in two respects. First,
it is now known\cite{r1} that the masses of any additional
neutrinos must be greater than $45$ GeV. It is thus quite possible that
the mass of the heavy charged lepton is degenerate with or smaller than
that of its neutral counterpart. In particular, the charged lepton is
mass degenerate with the heavy neutrino at lowest order
in models with vector-like leptons;
it can be lighter in models with right-handed singlet leptons.
These considerations lead to the possibility that the $L$ could be very
long-lived, perhaps not decaying inside a detector. Second, if the
right-handed $L$ and $N$ are in an $SU(2)$ doublet, the GIM mechanism
breaks down, leading to the flavor changing decay
 $L\rightarrow \tau Z$.

Discovery of such heavy leptons would revolutionize
our understanding of the fundamental fermion spectrum.
If they exist, it would be natural, by
consideration of quark-lepton symmetry, to expect further
quarks, beyond the top quark, to occur also, but in the
present article we shall not consider this possibility.

The layout of the present paper is as follows: Section II
discusses the four models containing heavy leptons; in
Section III are remarks on how detection of the heavy
leptons depends crucially on their lifetime which could
lie within a wide range, depending on the details
of the mass spectrum; the production cross section
formulae are
presented in Section IV; finally,
the results are provided in Section V.

\section{the models}
In the standard model, each of the three generations of
quarks and leptons mimics the first generation in
which the leptons transform under $SU(2)\times U(1)$ as
one doublet $(\nu_e,\ e^-)$ with $Y=-1\ (Q=T_3+\frac{1}{2}Y)$
and a singlet $e^+$ with $Y=+2$.

It is still unclear whether the $\nu _i\ (i=e,\ \mu ,\ \tau)\ $
are strictly massless or if there exist nonzero neutrino
masses. Evidence for the latter comes from at least two
sources: the solar neutrino measurements which suggest a
solar neutrino flux below that predicted by the standard
solar model\cite{r3}; the recent gallium experiment
results from SAGE\cite{r4} and GALLEX\cite{r5} lend
some support to the deficit established at the Davis
chlorine experiment\cite{r6} and at the Kamiokande
water detector\cite{r7}, suggesting neutrino
oscillations between massive neutrinos. A popular oscillation
 mechanism is that of  MSW\cite{r8} where the electron
neutrinos partially convert to muon neutrinos within the
interior of the Sun. Another evidence for a massive neutrino is
the $17$ keV neutrino claimed in the Simpson experiment
\cite{r9} and later experiments, but not reproduced in
other efforts\cite{r10}. All in all, none of these claims
clearly disproves that the first three neutrinos are
massless. On the other hand, we know from $Z^0$ decay
measurements\cite{r1} that any fourth neutrino coupling
normally to $Z^0$ must be heavier than $M_Z/2\sim 45$ GeV.

In our first model (model 1), we shall suppose that the
fourth-generation leptons fall into the following
representations\\
\begin{equation}
\left ( \begin{array}{l}
N\\
L
\end{array}\right )_L \;,\; L_R \;,\; N_R \;,
\end{equation}
similar to the three light families except for the inclusion
of the right-handed neutrino $N_R$ which allows a Dirac neutrino mass.

The second model (model 2)
will instead assume representations\\
\begin{equation}
\left ( \begin{array}{l}
N\\
L
\end{array}\right )_L\;,\;
\left( \begin{array}{l} N\\
L
\end{array}\right )_R \;.
\end{equation}
They are called {\it vector leptons} because both the left- and right-
handed components transform identically under $SU(2)_L$.

Our third and fourth models are inspired by
the aspon model\cite{r2} of CP violation. To solve
the strong CP problem, the aspon model incorporates
{\it vector quarks}
at a scale of a few hundred GeV. Only colored states
contribute to the relevant anomaly so that leptons
are not required in solving the strong CP problem but by
quark-lepton symmetry we may expect that such a
model possesses also vector leptons. The vector
quarks may be in $SU(2)$ doublets or singlets.
So there is  a corresponding choice
for the heavy leptons. Our third model (model 3)
will therefore contain vector lepton doublets as
in Eq. (2.2) above, appended to the aspon model of Ref. 2.
Finally, the fourth model(model 4) will contain singlets
\begin{equation}
L_L\;,\; N_L\;,\;  L_R\;,\; N_R\;,
\end{equation}
added to the aspon model with $SU(2)$-singlet vector quarks.

\section{detection}
In this section, we first note that $L$ could be very
long-lived. If it is lighter than the $N$, and if both $N$ and $L$
do not mix with the standard model leptons, then $L$ would be
absolutely stable. This would be a cosmological disaster;
cosmological and astrophysical arguments\cite{ra2} limit the lifetime
to under 100 years. In the models we are considering in this paper,
it is quite natural to have mixings, and thus the lifetime of $L$
is model dependent. Knowing the lifetime is crucial for
experimental detection: if it is under $10^{-13}$ seconds,
the $L$ will decay at the vertex; if it is between $10^{-13}$
and $10^{-8}$ seconds, it will decay in the middle of the
detector; if it is greater than $10^{-8}$ seconds, it will
pass through the detector, and will look like a muon.

Let us first consider model 1.
If the $N$ is heavier than the $L$, and if it does mix with a
lighter neutrino (taken to be
$\nu_{\tau}$), the $L$ lifetime will be increased by a
factor of $\sin^2\theta $ ( where $\theta $ is the
mixing angle) over the lifetime it would have if the $N$
were massless. For a $100$ GeV $L$, this gives a lifetime
 of O($10^{-21}$ sec.)/$\sin^ 2\theta $. What are plausible
values of $\sin^ 2 \theta $? In see-saw type models,
$\sin^ 2\theta$ is given by either $m_{\tau}/m_L$ or
by  $m_{\nu _{\tau }}/m_N $, depending on whether
the mixing can occur in the charged lepton sector or whether
it is confined
to the neutrino sector. In the former case, the lifetime is
O($10^{-19}$) seconds; i.e. $L$ will decay at the vertex.
However, in the latter case, the lifetime is O($10^{-11}$ sec.)
($10$ eV$/m_{\nu _{\tau}}$).  Therefore the lifetime is at least $10^{-12}$
seconds, and could easily be long enough that the $L$ would pass
through any detector.

In the case in which the $L_R$ and $N_R$ form a doublet (model 2
and model 3), the masses are degenerate at tree level. The $L$ and $N$
will acquire a mass splitting from radiative
corrections. This gives a splitting of $O(200)$ MeV;
the precise splitting depends on masses and on the particle
content of the model. This splitting gives a lifetime between
$10^{-12}$ seconds and $10^{-6}$ seconds, covering the entire
range of interest.

One can thus see that all three lifetimes: (a) decay at the vertex,
(b) decay in the detector and (c) decay outside the detector are all
plausible, and each possibility must be considered.

If $L$ decays before leaving the vertex, the analysis of the detection will
be the same as that for a conventional heavy lepton, with one
crucial exception.  The heavy $L$'s transform differently from the
standard model charged leptons in models 2, 3 and 4, and
the GIM mechanism will break down, leading to flavor changing decays
such as $L\rightarrow \tau Z$.

By neglecting the mass of $\tau$, one finds the ratio
(for $m_L >m_Z$):
\begin{equation}
\frac{\Gamma (L\rightarrow\tau Z)}{\Gamma (L\rightarrow \nu_{\tau} W)}=
\frac{|U_{L\tau}|^2}{2 cos^2\theta_W |U_{L\nu_{\tau}}|^2}\frac{(m_L^2-2m_Z^2
+m_L^4/m_Z^2)(m_L^2-m_Z^2)}{(m_L^2-2m_W^2+m_L^4/
m_W^2)(m_L^2-m_W^2)} \,.
\end{equation}
An estimate of the value of $U_{L\tau}$ can be made by analogy with
similar GIM violation in the aspon model\cite{r2} which gives
$U_{L\tau}=(m_{\tau}/m_L)x_{\tau}$, where $x_{\tau}$
gives the ratio of $M_{34}$ to $M_{44}$ in the lepton mass matrix.
$U_{L\nu_\tau}$ is expected to be of order of $\sqrt{m_{\tau}/m_L}$ or
$\sqrt{m_{\nu_{\tau}}/m_N}$. In the former case, one finds the branching ratio
to be of the order of a few percent; in the latter it is nearly one
hundred percent.
Even if we take a small branching ratio, the background for
a particle decaying into $Z \tau $ would be extremely small
(especially if a vertex detector could pick up the tau). A major
problem with the conventional heavy lepton detection
has been backgrounds;
the $L\rightarrow \tau Z $ signal, even with a branching
ratio as low as $1\%$, may be easy to pick up.

If the decay is in the middle of the detector, but away
from the vertex, it should be easy to detect. An apparent
muon will suddenly decay
into missing energy and a real or virtual $W$.
The backgrounds should be negligible.

If the decay is outside the detector, the $L$ will be
indistinguishable from a muon. The production cross section,
 as will be shown in the next section, is large enough that
 thousands of $L$'s could be produced annually at the SSC, but
 it is small compared with muon pair production, so the
``extra'' muons would not be noticed. One possible method
 of detection would be time-of-flight. Many of the
 $L$'s will have $\beta < 1\ ($ see
 Section V for $d\sigma /d\beta )$, and if timing is
installed in the detectors, the $L$'s could be seen.
It is interesting that $1000$'s of $L$'s
could be produced, but that they could be missed
 if timing is not present.

\section{production cross sections}
In this paper, we consider the production processes for
$pp\rightarrow L^+L^-$, $NN$, $NL^{\pm}$, as well as
$pp \rightarrow L^+L^-A$, $NNA $ and $NLA$ where $A$ is the aspon
in models 3 and 4.  The cross sections and Feynman graphs for all relevant
subprocesses are given in Appendix A and Figure 1 respectively.
The total cross sections for all the
above processes are computed by using EHLQ[13] parton structure
functions(set 1).

For model 1, gluon fusion production (see Fig. 1-a),
by $Z$ and $H$ exchange,
is more important because the cross sections are proportional to the
square of the lepton mass. For the vector lepton models
(models 2, 3 and 4),
gluon fusion will not contribute, since vector leptons do not couple to $H$
 and a vectorlike coupling to the $Z$ gives no contribution due to
Furry's theorem. Thus, the only contributions for the pair production
of leptons in these
models are by quark fusion (see Fig. 1-b) in which the
cross sections fall off faster.

In addition, an aspon A can be produced through the bremsstrahlung effect
from the heavy leptons (see Fig. 1-c).  For completeness, we include
also the production cross sections for
$pp \ra L^+L^-A$, $NNA $ and $NLA$ at the SSC and LHC.

As we discussed in the previous sections, a long-lived charged lepton
can only be discovered if there are timing facilities in detectors.
A long-lived charged lepton is more likely to appear in the vector lepton
doublet models such as models 2 and 3.
The velocity distribution, ${d\sigma/d\beta}$, where $\beta$ is
 defined as the ratio of the momentum to
the energy of $L$ in lab-frame, has been calculated at the LHC and SSC
energies for $m_L = 100, 300$ and $500$ GeV; the results are reported at
the end of the following section.

\section{results and conclusions}
The results for the production cross sections at the SSC ($\sqrt s=40$ TeV)
and the LHC ($\sqrt s=17$ TeV) are displayed for the different final states
 of $pp$ collisions in Figs. 2-6. From these figures one can estimate
easily the number of events per collider-year using the projected
luminosities of the two machines (SSC: $10^{33}$ cm$^{-2}$s$^{-1}$; LHC:
$10^{34}$ cm$^{-2}$s$^{-1}$) and the corresponding annual integrated
luminosities $10$ fb$^{-1}$ y$^{-1}$ and  $100$ fb$^{-1}$y$^{-1}$ respectively.

For heavy $L$ or $N$, the cross sections for
$pp\rightarrow L^+L^-$, $NN$ are largest for model 1 because of the
 dominant gluon fusion contribution
(with $Z$ and $H$ exchange) in which cross sections are
proportional to the square of the masses; there is no such contribution
for vector leptons (model 2, 3 and 4) because both the $Z$ and $H$
diagrams (Fig. 1-a) vanish, as discussed earlier.
In particular, for $pp\rightarrow L^+L^-$ (Fig. 2) and  $M_L=400$ GeV there
are predicted to be 10,000 events for model 1 per year at the SSC and
the LHC.  For models 2, 3 and 4 (where the gluon fusion contributions
vanish), there are 1,000 or 2,000 events for model 2 and 3; and
500 or 1,000 events for model 4 at the SSC or the LHC respectively.
Similar rates are predicted for $pp\rightarrow NN$ (which is not allowed
in model 4) although the photon contribution vanishes.
Finally, the cross sections for $pp\rightarrow NL^{\pm}$,
which are allowed by $W$ exchange, can be read off from Fig. 4.
Although the luminosity is ten times higher at the LHC, the number of
heavy leptons produced in general
is just two times that at the SSC.

Note that although the cross sections for model 2, 3 and 4 are
considerably smaller, these models do have an $L\rightarrow
\tau Z$ decay mode, and thus possibly a much cleaner signature, if
it decays in the detector.
For $pp\rightarrow NL^{\pm}$ in which only the $W$ exchange is allowed,
models 1, 2 and 3 give similar cross sections.

For $pp\rightarrow L^+L^-A$, $NNA$ and $NLA$ with an aspon in the
final state, the cross sections, which are shown in Figs. 5, 6 and 7
respectively, are about $100$ times smaller than
without an aspon, but are still within the range of detectability of
SSC and LHC. model 3 (heavy lepton doublets) gives a slightly larger
cross section than model 4 (heavy lepton singlets) because the former
allows certain $W$ and $Z$ couplings.

If timing facilities are installed in detectors, the $\beta$
distribution functions $1/\sigma ( d \sigma / d \beta)$ would be
relevant. In Fig. 8, we plot the $\beta$ distributions for
$pp\rightarrow L^+L^-$ in the vector doublet models (models 2 and 3) for
$m_L = 100$, $300$ and $500$ GeV at the LHC (Fig. 8-a) and the SSC
(Fig. 8-b).
For a muon, the distribution is, of course,
a delta function at $\beta = 1$; whereas the $\beta$ distribution
spreads out to $\beta<1$ for a heavy lepton with an enhancement near
$\beta=1$. From Fig. 8 we conclude that in searching
 for a long-lived charged lepton, time-of-flight
is a valuable method
because of the characteristic spreading to $\beta<1$; at SSC this is
viable up to at least $m_L=500$ GeV. Thus timing in the SSC detector
would be particularly useful.

\acknowledgments
One of us (Y. Y.) would like to thank the
Department of Physics and Astronomy of the University of North
Carolina at Chapel Hill for its hospitality,
and would also like to thank the Texas
National Laboratory Research Commission for support.
The work of P. H. F. was supported by the U.S. Department of
Energy under Grant No. DE-FG05-95ER-40219. The work of D. N. was
supported in part by the U.S. Department of Energy under Grant No.
DE-FG05-95ER-40219 and the Natural Sciences and Engineering Research
Council of Canada.  The work of M. S. was supported in
part by the National Science Foundation. We thank John Ellis and Ryczard
Stroynowski for useful discussions.

\newpage
\begin{appendix}
\\
The cross sections for the various subprocesses  are listed below.\\
\\[0.4cm]
\underline{ $gg\ra\lp\lm$ and $gg\ra NN$ }

This production mechanism, by Z and H exchange, is allowed in model 1
only.  The cross sections by Z and H exchange are given respectively by
\begin{eqnarray}
\hat {\s}_Z (gg\ra \lp \lm ) &=&  \f{\beta \alpha^2\alpha_s^2m_L^2}
              {2048\pi \sin^4\tw m_{\small W}^4}|I|^2 \;, \\
\label{eq:ZLL}
\hat {\s}_H (gg\ra \lp \lm ) &=& \f{\beta ^3\alpha ^2\alpha _s^2m_L^2}
                                 {4608\pi \sin^4\tw m_{\small W}^4}
                           \f{\she^2}{\hp }|J|^2 \;,
\label{eq:HLL}
\end{eqnarray}
where $\sqrt {\she}$ is the center of mass energy available for the
subprocess and $\beta $ defined as $\beta =\sqrt {1-4m_L^2/\she}$ $\;$ is
the velocity of $L$.  $I$ and $J$ are given by
\begin{eqnarray}
I&=&2\sum_q(\pm )\int_0^1 dx\int_0^{1-x} dy\f{xy}{xy-m_q^2/\she} \, ,\\
J&=&3\sum_q\int_0^1 dx\int_0^{1-x} dy\f{1-4xy}{1-xy\she/m_q^2}\, .
\end{eqnarray}
The sum runs over all known quarks and top-quark($m_t = 100$ GeV is
assumed). The $+(-)$ sign in the above equation applies to the quarks
with isopins $T_3 = 1/2(- 1/2)$.

$\hat {\s}_Z (gg\ra N N)$ and $\hat {\s}_H (gg\ra N N)$ are the same as
Eqs. (A1) and (A2)
respectively with $m_L$ replaced by $m_N$.
\\[0.4cm]
\underline {$q \bar q \ra\lp\lm$ and $q \bar q \ra NN$}

The cross section for  $q \bar q \ra\lp\lm$ (and $q \bar q \ra NN$,
see below) is given by
\begin{eqnarray}
\hat {\s} (\qll )& = &\f {2\pi \alpha ^2\beta B}{9\she}
\left(q_i^2-\f {q_i\she(\she-m_{\small Z}^2)(g_L^q+g_R^q)(g_L^l+g_R^l)}
{2\sin^2\tw \cos^2\tw \zp }  \right)  \nonumber \\
 & &\mbox{} +\f { \pi \alpha ^2\beta \she ({g_L^l}^2+{g_R^l}^2)
 (B(g_L^q+g_R^q)^2+2\beta ^2(g_L^q-g_R^q)^2)}
{36\sin^4\tw \cos^4\tw\zp }\;,
\end{eqnarray}
where $ B=3-\beta^2\ $ with $\beta =\sqrt{ 1 - 4 m_{L,N}^2/\she}$
, and $q_i e$ is the charge of the quark of type $i$.
$g_L^q=T_3-q_i\sin^2\tw$ and $g_R^q=-q_i\sin^2\tw$ are the quark and $Z$
boson neutral coupling coefficients.  For leptons, the coefficients
$g_L^l$ and $g_R^l$ for various models are given by
\begin{eqnarray}
g_L^l =
\left [
\matrix{
T_3 - Q_l \sin^2 \tw \;, & \rm{model\, 1,\,2 \,and \,3}\cr
-Q_l \sin^2 \tw \;, &   \rm{model \,4}}
\right. \; ,
\end{eqnarray}
and
\begin{eqnarray}
g_R^l =
\left [
\matrix{
T_3 - Q_l \sin^2 \tw \;, & \rm{model \,2 \,and \,3}\cr
-Q_l \sin^2 \tw \;,  & \rm{model \,1 \,and \,4}}
\right. \; ,
\end{eqnarray}
where $T_3 = 1/2 (- 1/2)$ and $Q_l = 0(-1)$ for $l = N (L)$.
For the process $qq\ra NN$, $q_i = 0$ is used in Eq. (A5)
because the photon does not contribute.
\\[0.4cm]
\underline {$q \bar q' \ra \ N L^{\pm} $}

The cross section of this sub-process is

\begin{equation}
\hat {\s} (q \bar q^{\prime} \ra NL^{\pm}) =
\f{\pi \alpha ^2|U_{qq\prime }|^2\beta \she F}{24\sin^4\tw\wpr } \, ,
\end{equation}\\[0.1cm]
where
\begin{eqnarray}
F =
\left [
\matrix{
0.5\left [ 1+\beta ^2/3-\left( (m_L^2-m_N^2)/\she \right)^2\right ] \;,
& \rm{model \, 1}\cr
\left [ 1+\beta ^2/3 -\left( (m_L^2-m_N^2)/\she\right)^2+3m_Lm_N/\she
\right ]\;,
& \rm{model \,2 \,and \,3}\cr
0,& \rm{model \, 4}
}
\right. \, ,
\end{eqnarray}
with $\beta=\left [1-2(m_L^2+m_N^2)/\she+((m_L^2-m_N^2)/\she
)^2\right ]^{1/2}$ is
again the speed of the charged lepton $L$ in the $q\bar q'$ center-of-mass.
\\[0.4cm]
\underline { $q \bar q \ra \lp \lm A$,  $q \bar q  \ra N N A$ and
             $q \bar q' \ra N L A$}

The amplitude squared of these sub-processes(in model 3 and 4 only), with
the momenta $p_1 + p_2 \ra p_3 + p_4 +p_5$ respectively, are given by
\begin{equation}
32({G_L}^2+{G_R}^2)(A_1+A_2+A_{12})
\;,
\end{equation}
with
\begin{eqnarray}
A_{1} &=&\f{1}{((p_3+p_5)^2-m_3^2)^2}
\nonumber\\
& & \times\left [(2p_3\cdot p_5-2m_3^2)(p_1\cdot p_4\;p_2\cdot p_5+p_1\cdot
p_5\;p_2\cdot p_4)
\right. \nonumber \\
& &  -(2m_3^2+m_A^2)(p_1\cdot p_3\;p_2\cdot p_4+p_1\cdot p_4\;p_2\cdot
p_3) \nonumber \\
& & \left. -2m_3m_4\;p_1\cdot p_2\;(m_3^2+m_A^2+p_5\cdot p_3)\right]
\;, \\
A_{2} &=& A_{1}(p_3\leftrightarrow p_4,\ m_3\leftrightarrow m_4) \;, \\
A_{12} &=& \f {1}{((p_3+p_5)^2-m_3^2)((p_4+p_5)^2-m_4^2)}\nonumber\\
& & \times\left[-4p_4\cdot p_5\;p_1\cdot p_3\;p_2
\cdot p_3-4\;p_3\cdot p_5\;
p_1\cdot p_4\;p_2\cdot p_4\right.  \nonumber \\
& &  +2p_3\cdot p_4\;(p_1\cdot p_4\;p_2\cdot p_5+p_1\cdot
 p_5\;p_2\cdot p_4+
p_1\cdot p_3\;p_2\cdot p_5\;+ p_1\cdot p_5\;p_2\cdot p_3) \nonumber \\
& & +(4p_3\cdot p_4+2p_4\cdot p_5+2p_3\cdot p_5)
 (p_1\cdot p_3\;p_2\cdot p_4+p_1\cdot p_4\;p_2\cdot p_3) \nonumber \\
& &+m_3m_4\;\left(-4p_1\cdot p_5\;p_2\cdot p_5
       +p_1\cdot p_2(2p_3\cdot p_5+2p_4\cdot p_5+4p_3\cdot p_4+2m_A^2)\right)
\nonumber\\
& & \left. +2m_A^2\;p_3\cdot p_4\;p_1\cdot p_2\right]
\end{eqnarray}
\vspace{-0.6cm}

In $A_1$, $A_2$ and $A_{12}$ the heavy lepton masses are taken to be
$m_3=m_4=m_L$ for $q \bar q \ra \lp \lm A$, $m_3=m_4=m_N$ for $
q \bar q \ra N N A$ and $m_3=m_N,m_4=m_L$ for $ q \bar q \ra N L A$.

Finally, in Eq. (A10), the values of $G_L$ and $G_R$ are given as follows:
\begin{equation}
\begin{array}{lcc}
 & G_L & G_R \\
q \bar q \ra L^+L^-A & \ \
g_A\left (g_L^Lg_L^qP-q_ie^2/\she\right )
& \ \ \   g_A\left (g_L^Lg_R
^qP-q_ie^2/\she\right )  \\
q \bar q \ra NNA &  g_A g_L^Ng_L^qP &  g_A g_L^Ng_R^qP\\
q \bar q \ra NLA\ (\rm{model\,3}) \  &
g_Ag^2 /2(\she-m_{\small W}^2) & 0\\
q \bar q \ra NLA\ (\rm{model\,4})& 0 & 0
\end{array}
\end{equation}
where  $P=\left (g/\cos\tw\right )^2/(\she-m_{\small Z}^2)$ and
 $g_L^L$ and $g_L^N$ in model 3 and 4 are given in
Eqs. (A6) for $l= L$ and $N$.
\end{appendix}

\newpage
\begin{center}FIGURE CAPTIONS
\end{center}

Fig.1.  Feynman diagrams for all the subprocesses.
(a) $g g \rightarrow L^{+}L^{-}$, and $ N N$;
(b) $q \bar q \rightarrow L^+L^-,\  N L^{\pm}$ and $N N$;
(c) $q \bar q \rightarrow L^+ L^- A,\  NLA$ and $N N A$.

Fig.2.  Total cross sections for heavy lepton production $p p\rightarrow
L^+L^-$ as a function of the charged lepton mass $m_L$ for model 1(solid
lines), model 2 and 3(dashed lines) and model 4(dotted lines).
The upper and lower sets are for $ \sqrt s = 40$ TeV  and $\sqrt s = 17$ TeV.
$m_H=100$ GeV is assumed.

Fig.3.  Total cross sections for the process $p p \rightarrow N N$
as a function of the heavy neutrino mass $m_N$ for model 1(solid lines),
and model 2 and 3 (dashed lines).
The upper and lower sets are  for $ \sqrt s = 40$ TeV  and $\sqrt s = 17$ TeV.
$m_H=100$ GeV is assumed.

Fig.4.  Total cross sections for the process $p p\rightarrow N L^{\pm}$
for model 1(solid lines), and model 2 and 3(dashed lines) for
(a) $m_N/m_L$ = 0.5,
(b) $m_N/m_L$ = 1,
(c) $m_N/m_L$ = 2.
The upper and lower sets are for $ \sqrt s = 40$ TeV  and $\sqrt s = 17$ TeV.
$m_H=100$ GeV is assumed.

Fig.5.  Total cross sections for the process $pp\rightarrow L^+L^-A(A=aspon)$
as a function of aspon mass $m_A$ for (a)model 3 and (b) model 4 with
$m_L = 50$ GeV(solid lines) and $m_L = 150$ GeV(dashed lines).
The upper and lower sets are for $ \sqrt s = 40$ TeV  and $\sqrt s = 17$ TeV.
$m_H=100$ GeV and the coupling of the aspon $\alpha_A=0.1$ are assumed.

Fig.6.  Total cross sections for the processes $pp\rightarrow NNA$
in model 3 for
$m_L = m_N = 50$ GeV(solid lines) and  $m_L = m_N = 150$ GeV(dashed lines).
The upper and lower sets are for $ \sqrt s = 40$ TeV  and $\sqrt s = 17$ TeV.
$m_H=100$ GeV and the coupling of the aspon $\alpha_A=0.1$ are assumed.

Fig.7.  Total cross sections for the processes $pp\rightarrow N LA$
in model 3 for
$m_L = m_N = 50$ GeV(solid lines) and  $m_L = m_N = 150$ GeV(dashed lines).
The upper and lower sets are for $ \sqrt s = 40$ TeV  and $\sqrt s = 17$ TeV.
$m_H=100$ GeV and the coupling of the aspon $\alpha_A=0.1$ are assumed.

Fig.8.  The velocity distributions $1/\sigma(d\sigma/d\beta)$
for the process $pp\rightarrow LL$ in model 2 and 3
at
(a)$\sqrt s=17$ TeV , and
(b)$\sqrt s=40$ TeV
for $m_L=100$ GeV(solid line), $300$ GeV(dashed line)
and $500$ GeV(dotted line) respectively.


\end{document}